%% file: main.tex
\documentclass[twocolumn]{aastex7}

\usepackage{amsmath}
\usepackage{enumitem}
\usepackage{xspace}
\usepackage{xcolor}

\undef\listofchanges
\usepackage[commandnameprefix=always,commentmarkup=uwave]{changes}
\definechangesauthor[color=cyan]{wf}

\input{macros.tex}
\newcommand{\CIPlusMinus}[1]{{#1[median]^{+#1[error plus]}_{-#1[error minus]}}}

\definecolor{mossgreen}{HTML}{088206}

\newcommand{\Msun}{\ensuremath{\rm{M}_{\odot}}}

\newcommand{\zetaeff}{\ensuremath{\zeta_{\rm{eff}}}\xspace}
\newcommand{\betaacc}{\ensuremath{\beta_{\rm{acc}}}\xspace}

\newcommand{\fmrr}{\ensuremath{f_{\rm{MRR}}}\xspace}

\newcommand{\qobs}{\ensuremath{q_{\rm{obs}}}\xspace}
\newcommand{\qbbh}{\ensuremath{q_{\rm{BBH}}}\xspace}

\newcommand{\zetaMRR}{\macros[zeta_eff_mrr_threshold]}

\definecolor{brightmaroon}{rgb}{0.7, 0.09, 0.09} 
\definecolor{purple}{rgb}{0.3, 0.2, 0.9} 

\begin{document}

\title{A Strongly Parametrized Mass Ratio Model for the Stable Mass Transfer Channel:\\ a Case Study of the 10\,\Msun\,Peak}

\author[orcid=0000-0002-2202-8640,sname='Godfrey']{Jaxen Godfrey}
\affiliation{University of Oregon, Institute of Fundamental Science}
\email[show]{jaxeng@uoregon.edu}

\author[orcid=0000-0001-5484-4987,sname='van Son']{Lieke A. C. van Son}
\affiliation{Department of Astrophysics/IMAPP, Radboud University Nijmegen, PO Box 9010, 6500 GL Nijmegen, The Netherlands}
\email[show]{lieke.vanson@ru.nl}  

\author[orcid=0000-0002-2916-9200, sname=Farr]{Ben Farr}
\affiliation{University of Oregon, Institute of Fundamental Science}
\email{bfarr@uoregon.edu}

\begin{abstract}
The mass ratio of merging binary black holes (BBHs) carries information about their formation history, yet has received less attention than masses, spins and eccentricities as a channel discriminator.
We derive a strongly parametrized analytical model for the mass-ratio distribution expected from the stable mass transfer (SMT) channel. 
The model maps mass-transfer stability and accretion efficiency onto the observed mass-ratio distribution, and naturally produces two qualitatively distinct subpopulations: 
a non-mass-ratio-reversed and a mass-ratio-reversed subpopulation whose distinct shapes depend on the binary-evolution parameters in a traceable way.
We embed this model in a hierarchical population analysis and apply it to the $\sim 10\Msun$ peak in the GWTC-4 BBH catalog. 
We find that the data favor little to no mass-ratio reversal in this peak, and infer SMT parameters in an astrophysically plausible range. 
This work demonstrates how data-driven models can be used in mixtures to study singular features in BBH population data and serves as a proof of concept for how a measurement of the BBH mass-ratio distribution within a subpopulation can be translated into direct constraints on the binary-evolution physics that produced it.
\end{abstract}


\input{intro.tex}

\input{method.tex}

\input{strong_results.tex}
\input{conclusion}

\begin{acknowledgments}
LvS is supported by VI.Veni.242.115, Grant ID \url{https://doi.org/10.61686/XVIAV86753}. This material is based upon work supported by the National Science Foundation Graduate Research Fellowship under Grant No. 2236419. This research has made use of data, software and/or web tools obtained from the Gravitational Wave Open Science Center 
(\url{https://www.gw-openscience.org/}), a service of LIGO Laboratory, the LIGO Scientific Collaboration and the Virgo Collaboration. 
BF acknowledges support from the National Science Foundation under grant PHY-2146528.
The authors are grateful for computational resources provided by the LIGO Laboratory and supported by National Science Foundation Grants PHY-0757058 and PHY-0823459.  
The authors acknowledge Research Advanced Computing Services (RACS) at the University of Oregon for providing computing resources that have contributed to the research results reported within this publication (\url{https://racs.uoregon.edu}). This material is based upon work supported 
in part by the National Science Foundation under Grant PHY-2146528 and work supported by NSF's LIGO Laboratory which is a major facility 
fully funded by the National Science Foundation.
\end{acknowledgments}

\software{
This work made use of the following software packages: \texttt{gwinferno} \citep{Edelman:2022ydv,Godfrey2023}, \texttt{astropy } \citep{astropy:2013,astropy:2018,astropy:2022,astropy_17756022}, \texttt{Jupyter } \citep{2007CSE.....9c..21P,kluyver2016jupyter}, \texttt{matplotlib } \citep{Hunter:2007}, \texttt{numpy } \citep{numpy}, \texttt{pandas } \citep{mckinney-proc-scipy-2010,pandas_19340003}, \texttt{python } \citep{python}, \texttt{scipy } \citep{2020SciPy-NMeth,scipy_18736568}, \texttt{ArviZ } \citep{arviz_2019,ArviZ_19723695}, \texttt{JAX } \citep{jax2018github}, \texttt{numpyro } \citep{phan2019composable,bingham2019pyro}, \texttt{seaborn } \citep{Waskom2021}, and \texttt{xarray } \citep{hoyer2017xarray,xarray_17834519}.  Software citation information aggregated using \texttt{\href{https://www.tomwagg.com/software-citation-station/}{The Software Citation Station}} \citep{software-citation-station-paper,software-citation-station-zenodo}.
This research has made use of the Astrophysics Data System, funded by NASA under Cooperative Agreement 80NSSC21M00561.
}







\bibliography{main}{}
\bibliographystyle{aasjournalv7}



\end{document}

%% file: intro.tex
\section{Introduction} \label{sec:intro}
Gravitational-wave (GW) observations promise a unique window into the physics of massive stellar binaries.
With each data release our ability to reconstruct the astrophysical population of GW sources has improved \citep{LIGO:GWTC1Pop, LIGO:GWTC2Pop, LIGO:GWTC3Pop, LIGO:GWTC4Pop}, yet connecting these observations to theoretical predictions has proven challenging; the physics of massive binary evolution involves a large, highly uncertain parameter space in which different assumptions can lead to vastly different compact object populations \citep{Belczynski:2021zaz, Agrawal:2023ecm, Iorio:2022sgz, Marchant:2023wno}.
%
Mass is the best-measured source property and has accordingly received the most attention, both in predicting and inferring the shape of the binary black hole (BBH) mass function \citep[e.g.,][]{Veske:2021qis, Tiwari:2021yvr, Edelman:2022ydv, LIGO:GWTC3Pop, Toubiana:2023egi, Ray:2024hos, CallisterFarr2024, LIGO:GWTC4Pop}.
Yet there is significant overlap between the mass distributions predicted by different formation channels \citep{Mandel:2018hfr,Mandel:2021smh,2021hgwa.bookE..16M}, making it difficult to disentangle the astrophysical origin of the observed BBH population from mass alone.
%
Considerable effort has therefore gone into using (effective) spins \citep[e.g.,][]{2016ApJ...832L...2R,2017CQGra..34cLT01V,2017Natur.548..426F,2018ApJ...854L...9F} and eccentricity \citep[e.g.,][]{2018PhRvD..98h3028L,2019MNRAS.490.5210R,2019ApJ...871...91Z,2021ApJ...921L..31R,2025ApJ...994L..47S} as formation-channel discriminators, and more recently into identifying distinct BBH subpopulations through their joint mass, mass-ratio, and spin distributions \citep[e.g.,][]{Godfrey2023,Banagiri2025,2025CQGra..42v5008K,2025arXiv251022698W,2025arXiv250904637A,2025arXiv251105316T,2025arXiv251122093S}.
By comparison, the \textit{mass-ratio} distribution remains relatively understudied.
%
In GW population inference, the mass-ratio distribution is typically modelled as a single power law, often inferred as a function of primary mass.
In studies that assume both BH components are drawn from a single mass distribution, the mass-ratio distribution simply accounts for deviations from random pairing.
Models beyond a single power law have been less explored, with notable exceptions adopting flexible non-parametric forms \citep{2017MNRAS.465.3254M,2021ApJ...913L..19T, Edelman:2022ydv, Godfrey2023, CallisterFarr2024, Sadiq:2023zee, Heinzel2025, Colloms:2025qzu}; several of these studies hint that different parts of the mass spectrum prefer different mass-ratio distributions \citep{Baibhav:2022qxm, Li:2022jge, Godfrey2023, Sadiq:2023zee, 2025A&A...694A.186G, 2025CQGra..42v5008K}.

Yet the mass-ratio distribution is potentially a powerful channel discriminator, with distinct predictions across formation pathways.
Dynamical formation can produce a wide range of mass ratios but most predictions peak near unity \citep{Rodriguez:2016kxx, Torniamenti:2024uxl}, and chemically homogeneous evolution likewise predicts a strong preference for equal masses due to mass transfer during the contact phase \citep{deMink:2016, Marchant2016}.
Stable mass transfer (SMT) in isolated binaries, on the other hand, places very particular constraints on the final mass ratio of systems formed from two SMT events \citep[e.g.,][]{vanSon2022_nopeaks}, producing a distribution that peaks away from equal masses, near $q \sim 0.7$ (see Fig.~9 in \citealt{vanSon2022_rateRedshift}).
Notably, the GWTC-4.0 population analysis reports $q = 0.74^{+0.13}_{-0.13}$ for systems in the $\sim 10\,\Msun$ peak \citep{LIGO:GWTC4Pop}, already favouring non-equal-mass pairing.

The $\sim 10\,\Msun$ peak is therefore a particularly compelling starting point for investigating the mass-ratio distribution.
This feature is robustly recovered across a wide range of inference methods \citep{LIGO:GWTC3Pop, Edelman:2022ydv,Godfrey2023,CallisterFarr2024,Afroz:2024fzp, Afroz:2025ikg, LIGO:GWTC4Pop, Banagiri2025, 2026arXiv260401420L}.
Moreover, low-mass BBHs are likely dominated by isolated binary evolution \citep[e.g.,][]{2022MNRAS.517.4034S}, reducing the number of competing channels to disentangl, though some dynamical formation scenarios can produce a peak at low masses \citep[e.g.,][]{Torniamenti:2024uxl, Ye:2025ano}.
The SMT channel in particular has been proposed as a natural explanation for this peak: \citet{vanSon2022_nopeaks} showed that the global peak near $10\,\Msun$ may emerge naturally from mass-transfer stability constraints, without invoking a remnant-mass function that prefers this mass.
Alternative explanations attribute the peak to compactness-driven remnant-mass features \citep[e.g.,][]{2023ApJ...950L...9S,2025arXiv250820787W,2025arXiv251007573W,2025A&A...694A.186G,2026arXiv260401420L}, but even in these scenarios the BHs must be brought together by some pairing process, and the mass-ratio distribution retains diagnostic power for distinguishing between formation pathways.

Motivated by these considerations, we build on the analytical framework of \citet{vanSon2022_nopeaks} to derive a strongly parametrized model for the mass-ratio distribution expected from the SMT channel, with explicit dependence on the binary-evolution parameters.
We find that this model can naturally produce two qualitatively distinct subpopulations for systems that have not undergone mass-ratio reversal (non-MRR), and those that have (MRR).
The relative shapes of these two subpopulations depend on the SMT channel parameters. 
In the regime that we find is most relevant to the $\sim 10\,\Msun$ peak, the non-MRR subpopulation is broad and inherits the shape of the underlying ZAMS mass-ratio distribution (extending to low $q$), while the MRR subpopulation is narrower and concentrated near $q = 1$. 
Other parameter combinations can flip this picture, produce only MRR systems, or produce no MRR systems at all.
We embed this model in a hierarchical population analysis following \citet{Godfrey2023} and apply it to the $\sim 10\,\Msun$ peak in GWTC-4.0.
As a proof of concept, we show how a measurement of the mass-ratio distribution within a subpopulation of the BBH catalog can be translated, through this model, into a direct constraint on the binary-evolution physics that produced it.

The paper is structured as follows.
Section~\ref{sec: method} presents our analytical SMT mass-ratio model and the hierarchical inference framework we use to fit it to GW data.
Section~\ref{sec:results} presents the results of applying this analysis to the $10\,\Msun$ peak in GWTC-4.0.
We discuss our findings and outline directions for future work in Section~\ref{sec:conclusion}.

%% file: method.tex
\section{Method} \label{sec: method}
We first derive a strongly parametrized mass ratio model inspired by the stable mass transfer channel in Section~\ref{ss:smt q model}. 
In Section~\ref{ss: hierarchical model} we describe the hierarchical inference model used to fit the mass ratio distribution of the low-mass peak in the primary mass distribution. The dataset used is described in Section~\ref{ss: data}.

\subsection{A mass ratio model inspired by the stable mass transfer channel} \label{ss:smt q model}

We broadly follow the cartoon-like analytical approximation for the stable mass transfer (SMT) channel from \citet{vanSon2022_nopeaks}, which we briefly summarize here.
In this approximation, mass transfer (MT) between donor and accretor is dynamically stable when the donor-to-accretor mass ratio at the onset of MT, $q_\text{D,A}$, lies below a critical value $q_\text{crit}$. 
Throughout this work, we use subscripts $a$ and $b$ to denote the components that evolved from the more massive and less massive stars at Zero Age Main Sequence (ZAMS), respectively. 
Each MT event has its own critical value; i.e., $q_\text{crit,1}$ for MT1 from star $a$ to star $b$. Similarly, $q_\text{crit,2}$ for MT2 from star $b$ to star $a$.
Only systems that satisfy both conditions are considered to originate from the SMT channel.

After each MT event, a fraction $f_\text{core}$ of the donor's pre-MT mass remains as a bare (helium) core, and a fraction $\beta_\text{acc}$ of the mass lost by the donor is accreted by its companion. 
The core subsequently collapses to a BH via supernova (SN), losing a fraction $f_\text{SN}$ of its pre-SN mass in the process. Unlike \citet{vanSon2022_nopeaks}, we parameterize SN mass loss as a fraction rather than an absolute amount; this makes the final model independent of mass and yields bounds only on the allowed mass ratios, with no associated mass minimum.

Stability requires $\zeta_\text{RL} \leq \zetaeff$, where $\zeta_\text{RL}$ is the response of the Roche radius to mass loss and $\zetaeff$ is the adiabatic response of the donor \citep[cf.][]{Soberman:1997mq}. 
$\zeta_\text{RL}$ can be expressed in terms of $\beta_\text{acc}$ and $q_\text{D,A} = M_D/M_A$, so for given $\beta_\text{acc}$ and $\zetaeff$ the stability condition becomes a limit on the donor-to-accretor mass ratio,
\begin{equation}
    q_\text{D,A} \leq q_\text{crit}(\beta_\text{acc}, \zetaeff).
\end{equation}
Note that our formulation treats $\zetaeff$ and $\betaacc$ as single population-wide values, an assumption we briefly discuss further in Section~\ref{sec:conclusion}.

The initially more massive star ($a$) fills its Roche lobe first and triggers MT1; approximating both stars' pre-MT1 masses by their ZAMS values (i.e., ignoring wind-mass loss), MT1 is stable if
\begin{equation} \label{eq:crit1}
    q^1_\text{D,A} = \frac{M_\text{ZAMS,a}}{M_\text{ZAMS,b}} \leq q_\text{crit,1}(\beta_\text{acc}, \zetaeff).
\end{equation}
After MT1, star $b$ has mass $M_\text{b,postMT1} \geq M_\text{ZAMS,b}$ and star $a$ eventually collapses to a BH of mass $M_\text{BH,a}$. These are the component masses at the onset of MT2, which is stable if
\begin{equation} \label{eq:crit2}
    q^2_\text{D,A} = \frac{M_\text{b,postMT1}}{M_\text{BH,a}} \leq q_\text{crit,2}(\beta_\text{acc} = 0, \zetaeff).
\end{equation}
We set $\beta_\text{acc} = 0$ for MT2 assuming Eddington limited accretion onto the black hole. 

The constraints in \eqref{eq:crit1} and \eqref{eq:crit2} can now be recast as a range $(q_\text{BBH,min}, q_\text{BBH,max})$ on the final BH mass ratio $q_\text{BBH} = M_\text{BH,b}/M_\text{BH,a}$, with
%
\begin{equation} \label{eq:bbh_min}
    q_\text{BBH,min} = \frac{1 - f_\text{SN,b}}{1-f_\text{SN,a}}\Big[ \frac{1}{q_\text{crit,1}} + \beta_\text{acc}(1-f_\text{core})\Big],
\end{equation}
\begin{equation} \label{eq:bbh_max}
    q_\text{BBH,max} = q_\text{crit,2}\,f_\text{core}(1 - f_\text{SN,b}),
\end{equation}
where we take $f_\text{core}$ to be the same for both components\footnote{Note however that $f_\text{core}$ will in reality depend on the mass of the star.}, and $f_\text{SN,a}$, $f_\text{SN,b}$ are the SN mass-loss fractions. Equivalently, the mass ratio of black hole $b$ over black hole $a$ can be expressed directly in terms of the ZAMS mass ratio;
\begin{equation}\label{eq:q_bbh}
    q_\text{BBH} = \frac{1 - f_\text{SN,b}}{1-f_\text{SN,a}}\Big[ q_\text{ZAMS} + \beta_\text{acc}(1-f_\text{core})\Big].
\end{equation}

This analytical prescription connects the final BBH system to its progenitor in terms of the SMT channel parameters. 
To connect it to observations, we must also account for mass-ratio reversal. 
Although $q_\text{ZAMS} = M_\text{ZAMS,b}/M_\text{ZAMS,a} \leq 1$ by definition, $q_\text{BBH} \leq 1$ is not guaranteed: 
the BH descended from the initially less massive star can end up heavier than its companion. 
When we observe a BBH with GWs, we do not know if it underwent mass ratio reversal.
Hence for an observed mass ratio $q_\text{obs} = m_2/m_1 \leq 1$ there are two possible underlying values: $q_\text{BBH} = q_\text{obs}$ (no reversal) or $q_\text{BBH} = 1/q_\text{obs}$ (reversal).

\begin{figure*}
    \centering
    \includegraphics[width=1. \textwidth]{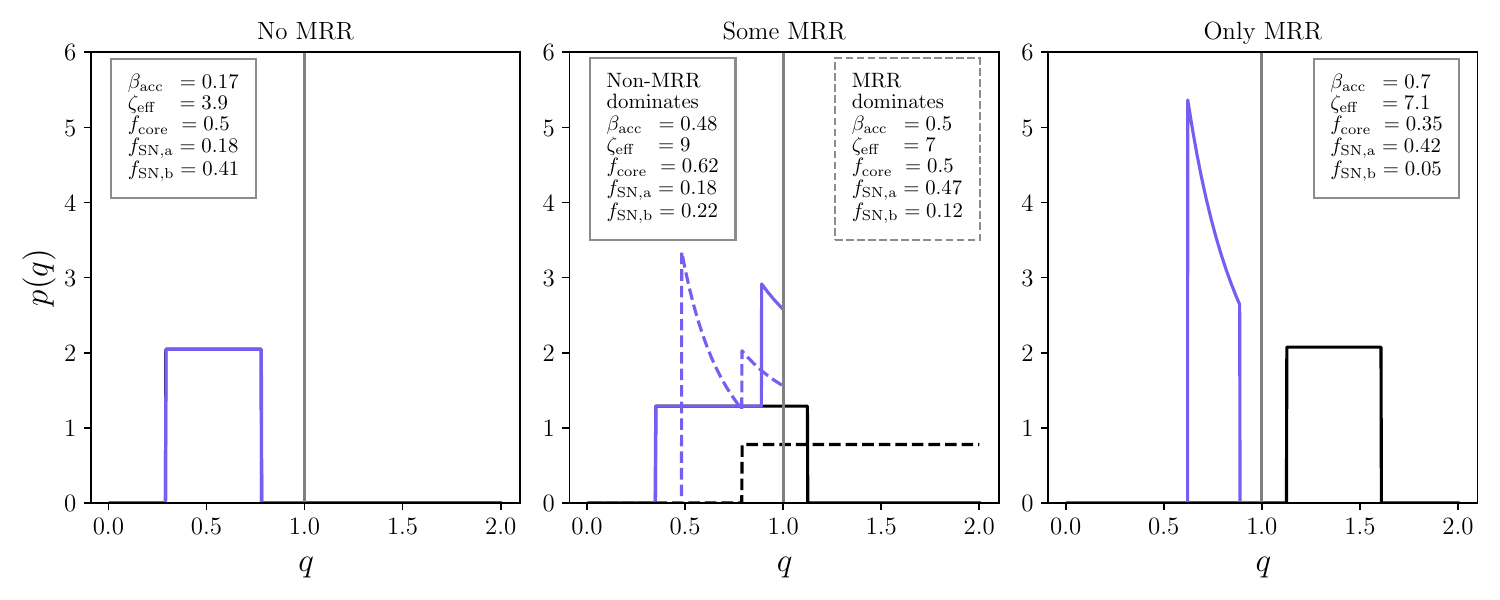}
    \caption{Example intrinsic (\qbbh, eq. \ref{eq:q_bbh} black) and observed (\qobs, eq. \ref{eq: q obs} purple) mass-ratio distributions predicted by our analytical SMT model for a uniform $q_\text{ZAMS}$ distribution. The possible morphologies are shown for each parameter regime: no MRR systems (left), some MRR systems (middle), and only MRR systems (right). In the left panel, the \qobs and \qbbh distributions are identical, as \qbbh $=$ \qobs in this regime.
    In the middle panel, the solid lines reflect an example distribution when non-MRR systems dominate over MRR systems, while the dashed lines reflect a distribution when MRR systems dominate. 
    In the former case, Mass-ratio reversal produces a pile-up of \qobs near equal masses, and the \qbbh--\qobs Jacobian shifts the peak away from $q = 1$. 
    In the right panel, \qbbh $>1$ which means $q_{\rm BBH, max}$ now maps to $q_{\rm obs, min}$ and the \qbbh--\qobs Jacobian produces a sharp peak. 
    An interactive version of this figure is available at \href{here}{https://gw.observer/research/smt-widget}.}
    
    \label{fig: analytical q}
\end{figure*}

Using \eqref{eq:q_bbh}, we can write the model for $q_\text{obs}$ as a function of the underlying $q_\text{ZAMS}$ distribution and the SMT channel parameters $\Lambda_\text{SMT} = \{\zetaeff, \beta_\text{acc}, f_\text{core}, f_\text{SN,a}, f_\text{SN,b}\}$,
\begin{align} \label{eq: q obs}
    p(q_{\rm obs}|\vec{\Lambda}_{\rm SMT}) = \frac{1}{C} & \frac{(1-f_{\rm SN,a})}{(1-f_{\rm SN,b})} \times \\ \nonumber
    \Bigg[ p(q_{\rm ZAMS}(q_{\rm obs})|&\vec{\Lambda}_{\rm SMT}) 
    + p\Big(q_{\rm ZAMS}\big(\tfrac{1}{q_{\rm obs}}\big)|\vec{\Lambda}_{\rm SMT}\Big) \frac{1}{q_{\rm obs}^2} \Bigg],
\end{align}
where the second term accounts for mass-ratio-reversed systems and $1/C$ is the normalization factor (computed numerically). The $q_\text{BBH}$ limits in \eqref{eq:bbh_min} and \eqref{eq:bbh_max} translate via \eqref{eq:q_bbh} into limits on $q_\text{ZAMS}$, and we adopt a uniform $q_\text{ZAMS}$ distribution \citep[e.g.,][]{2012Sci...337..444S} truncated at these values,
\begin{equation}\label{q_zams_dist}
    p(q_\text{ZAMS}|\Lambda_\text{SMT}) = \mathcal{U}(q_\text{ZAMS,min}, q_\text{ZAMS,max}).
\end{equation}

The SMT parameters control where the limits $q_\text{BBH,min}$ and $q_\text{BBH,max}$ fall, and hence the shape of the resulting $q_\text{obs}$ distribution. 
Of the five SMT parameters, $\zetaeff$ plays a particularly important role: larger $\zetaeff$ values widen the range of stable-MT mass ratios and so make it easier to produce BBHs via this channel. From equations~\eqref{eq:bbh_min} and \eqref{eq:bbh_max}, $q_\text{BBH,min}$ depends on all five SMT parameters. 
Though \eqref{eq:bbh_max} does not explicity depend on $f_{\rm SN,a}$, $q_\text{BBH,max}$ will depend on $f_{\rm SN,a}$ for parameter combinations that lead to $q_{\rm ZAMS,max} > 1$ because we define $q_{\rm ZAMS} \leq 1$. In this scenario, $q_{\rm BBH, max} = (1 - f_{\rm SN, b})/(1 - f_{\rm SN,a})\Big[ 1 + \beta(1 - f_{\rm core})\Big]$. In all other cases $q_{\rm BBH, max}$ depends on all SMT parameters except for $f_{\rm SN,a}$.

Figure \ref{fig: analytical q} shows the resulting $q_\text{BBH}$ and $q_\text{obs}$ distributions for example sets of SMT parameters with \eqref{q_zams_dist} as the $q_{\rm ZAMS}$ distribution. 
There are four different parameter regimes that correspond to four unique morphologies, defined by where the \qbbh bounds lie relative to $q = 1$. When $q_{\rm BBH, max} < 1$ and $q_{\rm BBH, min} < 1$, there are no mass-ratio reversed systems in the population, so the \qobs and \qbbh distributions are identical and exhibit a flat plateau between $q_{\rm BBH,min}$ and $q_{\rm BBH,max}$ (left-most panel in Figure \ref{fig: analytical q}).

When $q_{\rm BBH, min} < 1$ and $q_{\rm BBH, max} > 1$, the population contains a mixture of non-reversed and reversed systems. The relative porportion of these two branches is determined by the separation between $q_{\rm BBH} = 1$ and $q_{\rm BBH,min}$ and $q_{\rm BBH,max}$. The non-reversed branch maps $q_{\rm BBH} < 1$ directly to \qobs, while the reversed portion with $q_{\rm BBH} > 1$ is folded onto $q_{\rm obs} \leq 1$ via $q_{\rm obs} = 1/q_{\rm BBH}$ and reweighted by the Jacobian $1/q_{\rm obs}^2$. 
Two effects therefore shape the relative appearance of the two subpopulations: 
(i) the relative \emph{width} in $q_{\rm BBH}$-space of the non-reversed and reversed branches, 
set by where $q_{\rm BBH} = 1$ falls between $q_{\rm BBH,min}$ and $q_{\rm BBH,max}$, 
and (ii) the $1/q_{\rm obs}^2$ Jacobian, which redistributes the reversed branch toward smaller 
$q_{\rm obs}$, placing its peak at $q_{\rm obs} = 1/q_{\rm BBH,max}$ (the lower edge of the reversed interval).
For populations with fewer MRR systems than non-MRR systems (solid line in center panel of Figure~\ref{fig: analytical q}), the reversed systems pile up between $1/q_{\rm BBH,max} \leq q_\text{obs} \leq 1$, while the non-reversed systems extend down to $q_{\rm obs, min}$. 
The MRR peak tends to be a narrow feature near $q_{\rm obs} = 1$, while the non-MRR component tends to be broader, with more support at $q_{\rm obs} \sim 0.5$. When MRR systems dominate over non-MRR systems (the dashed line in the center panel of Figure~\ref{fig: analytical q}), $q_{\rm obs, min} = 1/q_{\rm BBH,max}$ corresponds to a sharp MRR peak that decays toward $q_{\rm obs} = 1$ and a smaller, secondary non-MRR peak at $q_{\rm obs} = q_{\rm BBH,max}$ that also
decays as \qobs approaches 1. The MRR peak tends to sit well below equal masses, while the non-MRR peak is squeezed into a range below just below $q_{\rm obs} = 1$.

Finally, when $q_{\rm BBH, min} > 1$ and $q_{\rm BBH, max} > 1$, the population only contains mass-ratio reversed systems (right-most panel of Figure \ref{fig: analytical q}). In this case, the Jacobian factor causes the observed distribution to peak at $q_{\rm obs,min} = 1/q_{\rm BBH, max}$ and decay towards $q_{\rm obs, max} = 1 / q_{\rm BBH, min}$.\footnote{We encourage the reader to use the interactive version of Figure~\ref{fig: analytical q} (\href{here}{https://gw.observer/research/smt-widget}) to explore how the two subpopulations trade off across the SMT parameter space.}

Qualitatively, the parameters that most impact the proportion of MRR to non-MRR systems in the mixed case are $f_{\rm SN,a}$ and $f_{\rm SN,b}$. When $f_{\rm SN, a} \approx f_{\rm SN,b}$, non-MRR systems are likely to dominate the population; when $f_{\rm SN,a} \gtrapprox f_{\rm SN,b}$, MRR systems are more likely to dominate. This is because more mass lost by the initially more massive binary component increases the chances that the intially less massive binary component will end up as the heavier BH. If $f_{\rm SN,a} < f_{\rm SN,b}$, there may be no MRR systems, whereas if $f_{\rm SN,a} \gg f_{\rm SN,b}$, only MRR systems will remain.

\subsection{Hierarchical inference model} \label{ss: hierarchical model}

\begin{figure*}
    \centering
    \includegraphics[width=0.9 \textwidth]{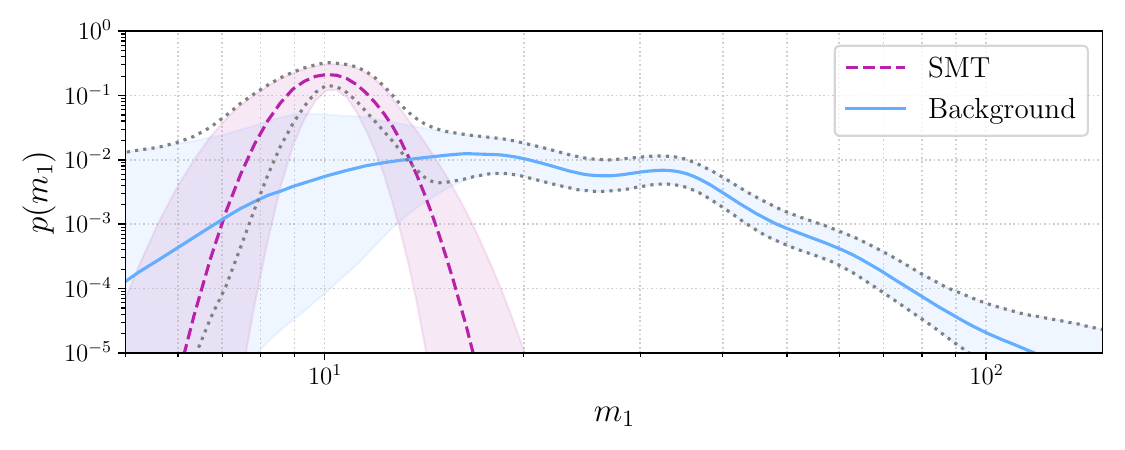}
    \caption{Inferred primary mass distribution under our model described in Section~\ref{ss: hierarchical model}. The log-Gaussian foreground component captures the $\sim 10\,\Msun$ peak, while the B-spline background captures the rest of the distribution. Shading indicates the 90\% credible interval. Dotted lines show the combined population. }
    \label{fig:primary masses}
\end{figure*}

Our approach is similar to the Isolated Peak model of \citet{Godfrey2023} and \citet{LIGO:GWTC4Pop}: we model the observed BBH catalog as a mixture of a foreground subpopulation, represented by a log-Gaussian peak in primary mass, and a background subpopulation, represented by a Basis-Spline (B-spline) in primary mass. 
The inferred primary mass distribution is shown in Figure \ref{fig:primary masses}.

To target the $\sim 10\,\Msun$ peak, we restrict the mean of the foreground log-Gaussian to $\mu \in [7,15]\,\Msun$. 
Note that this restriction does not force the foreground component to dominate in this mass range: the foreground branching fraction $\lambda_{\rm P1}$ is free to go to zero, in which case the B-spline background would account for the full distribution.
As in \cite{Godfrey2023}, the two subpopulations share a redshift distribution, modelled as a power law modulated by a B-spline, but each is associated with its own B-spline spin distribution.
The mass ratio distribution of the background is also inferred with a B-spline. 
The key difference from Isolated Peak lies in the mass ratio distribution of the log-Gaussian peak: rather than a B-spline, we adopt the analytical SMT prescription derived in Section~\ref{ss:smt q model}, given by equation~\eqref{eq: q obs}. 

The full mixture model is now given by
\begin{align}
p_{S}(\vec{\theta}|\vec{\Lambda}) = \, & \lambda_{\rm P}\, p_{\rm P}(\vec{\theta}|\vec{\Lambda}_{\rm P}) \\ \nonumber
& + (1 - \lambda_{\rm P})\, p_{\rm C}(\vec{\theta}|\vec{\Lambda}_{\rm C}),
\end{align}
where $\vec{\theta} = \{m_1, q_{\rm obs}, a, \cos t\}$ are the observed BBH parameters and $\vec{\Lambda}$ are the model hyper-parameters. The foreground (low-mass peak) component is
\begin{align}
    p_{\rm P}(\vec{\theta}|\vec{\Lambda}_{\rm P}) = \mathcal{N}_{\ln}(m_1|\mu,\sigma)\, p(q_{\rm obs}|\vec{\Lambda}_{\rm SMT})\, \mathcal{B}(a, \cos t|\vec{c}_{\rm P}),
\end{align}
where $\lambda_{\rm P}$ is the astrophysical branching fraction of this component. The background (continuum) component is
\begin{equation}
    p_{\rm C}(\vec{\theta}|\vec{\Lambda}_{\rm C}) = \mathcal{B}(\ln m_1, q, a, \cos t|\vec{c}_{\rm C}).
\end{equation}
We use the notation $\mathcal{N}_{\ln}(\cdot|\mu,\sigma)$ to denote a log-Gaussian and $\mathcal{B}(\cdot|\vec{c})$ a B-spline density with coefficients $\vec{c}$. 
The redshift distribution, common to both components, is suppressed in the notation above.

In this work, we allow $\zetaeff$ and $\betaacc$ to be the free parameters of our model with priors $\zeta_{\rm eff} \sim \mathcal{U}(3,7)$ and $\beta_{\rm acc} \sim \mathcal{U}(0,1)$ while we fix the remaining SMT parameters to reasonable values, $f_{\rm SN,a} = f_{\rm SN,b} = 0.2$ and $f_{\rm core} = 0.34$. We use a uniform prior for the astrophysical branching fraction, $\lambda_{\rm P} \sim \mathcal{U}(0,1)$. The priors for the B-spline coefficients are the same as those used in \citet{LIGO:GWTC4Pop}.

\subsection{Dataset and likelihood} \label{ss: data}

We use the BBH merger sample from the fourth Gravitational-Wave Transient Catalog \citep[GWTC-4.0;][]{LIGO:GWTC4}. We adopt the same waveform models and false-alarm-rate (FAR) threshold as the GWTC-4.0 population analysis \citep{LIGO:GWTC4Pop}, retaining the 153 BBH events that pass these cuts. 
Single-event posterior samples and the associated injection set \citep{Essick:2025zed} used to compute selection effects are taken from the public GWTC-4.0 data release.

We sample the posterior of our hierarchical population model using the \texttt{NUTS} \citep{1111.4246} sampler within the probabilistic programming language \texttt{NumPyro} \citep{phan2019composable,bingham2019pyro}. We use the same hierarchical likelihood as in \citet{LIGO:GWTC3Pop}, which involves Monte-Carlo summations over each single-event posterior, and apply the same prior constraint on the number of effective samples $N_{\rm eff}$ used in the Monte-Carlo summations to mitigate statistical error in the likelihood estimators.


%% file: strong_results.tex
\section{Results}\label{sec:results}

Figure~\ref{fig:primary masses} shows the inferred primary mass distribution for the foreground (SMT, purple) and background (blue) sub-populations.
%
In the foreground sub-population, the distribution peaks at $m_1 = \CIPlusMinus{\macros[peak_location]}\,\Msun$, in agreement with the low-mass peak that has now been identified across a wide range of population studies using a variety of statistical approaches \citep[e.g.,][]{Tiwari:2021yvr,2022ApJ...924..101E,Godfrey2023,Farah2023,CallisterFarr2024, Afroz:2024fzp, Afroz:2025ikg}, and consistent with the location reported in the O4a population analysis \citep{LIGO:GWTC4Pop}.
The background sub-population captures the rest of the distribution, including the secondary peak near $m_1 \approx 35\,\Msun$ and the tail at higher masses. 
The background does extend into the low-mass regime, but in this region its inferred rate density is roughly an order of magnitude below that of the Gaussian foreground component, and its uncertainty is large: the median background contribution at $m_1 \sim 10\,\Msun$ is well below the Gaussian, while the upper edge of the 90\% credible band reaches $p(m_1) \sim 0.05$, still a factor of two to four below the foreground peak. 
The relative heights of the foreground and background subpopulations indicates the proportion of events that contribute to each sub-population. Our mixture model differentiates the events in this mass regime using the properties of the other parameters, i.e. mass ratio and spin, which in turn contributes to the relative height of the two mass distributions. The uncertainty in the background mass distribution indicates that there may be a non-zero fraction of events in the $\sim 10 \, \Msun$ regime that have mass-ratio and spin properties that are more consistent with the background, and therefore potentially the product of a different channel, such as dynamical assembly, but we cannot determine confidently if this is the case. 

\begin{figure*}
    \centering
    \includegraphics[width=0.8\textwidth]{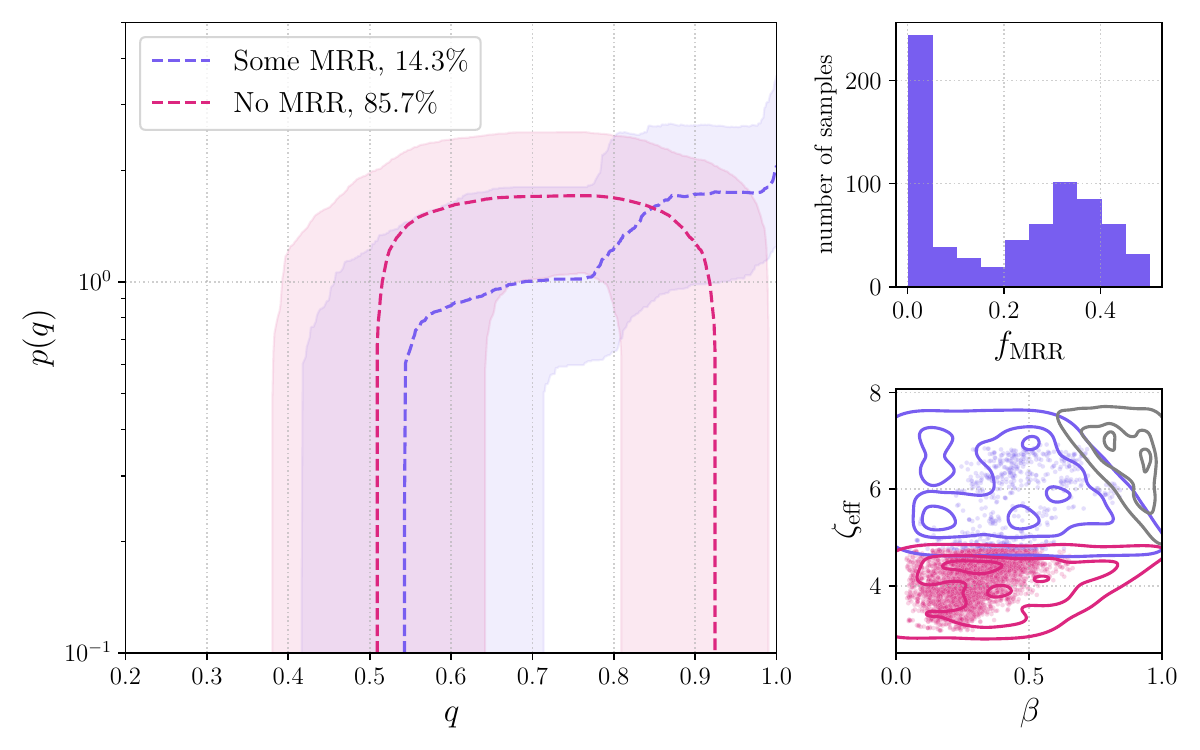}
    \caption{
    Inferred mass-ratio distribution for systems in the $\sim 10\,\Msun$ peak under our strongly parametrized model.
    \textbf{Left:} Inferred mass-ratio distribution, split between hyper-posterior samples with no mass-ratio reversal (no MRR; pink, $\macros[no_mrr_posterior_percent]\%$) and some mass-ratio reversal (some MRR; purple, $\macros[mrr_posterior_percent]\%$). Shading indicates the 90\% credible interval.
    \textbf{Top right:} Posterior on the MRR fraction ($\fmrr$, eq. \ref{eq: fmrr}) for ``some MRR'' samples, showing support for a small MRR fraction.
    \textbf{Bottom right:} 
    Joint posterior samples on SMT parameters $\zeta_{\rm eff}$ and $\beta_{\rm acc}$, split by presence (purple) or absence (pink) of MRR. Contours show prior bounds for three scenarios: no MRR (pink), non-MRR dominant (purple), and MRR dominant (gray).
    Overall, the data prefer little to no mass-ratio reversal in the $10\,\Msun$ peak, with SMT parameters in an astrophysically reasonable range.
    }
    \label{fig:q strong model}
\end{figure*}

We infer the mass ratio distribution of the $\sim 10\,\Msun$ peak using the SMT model from equation~\eqref{eq: q obs}; the results are summarized in Figure~\ref{fig:q strong model}.
Because $q_{\rm BBH,max}$ depends only on $\zeta_{\rm eff}$ (equation~\ref{eq:bbh_max}), MRR is only possible when $\zeta_{\rm eff} > \zetaMRR$. We can therefore split the hyper-posterior into two regions: one in which MRR is allowed ($\zeta_{\rm eff} > \zetaMRR$)
, and one in which it is forbidden ($\zeta_{\rm eff} < \zetaMRR$)

The left panel of Figure \ref{fig:q strong model} shows the inferred mass ratio distribution for the two branches of the posterior. 
We use dark purple to display the portion of the hyper-posterior where some MRR systems are possible, which accounts for $\macros[mrr_posterior_percent]\%$ of the total hyper-posterior. 
For these samples, we additionally compute the fraction of systems in the subpopulation that have undergone MRR,
\begin{equation}
    f_{\rm MRR} = \frac{q_{\rm BBH,max} - 1}{q_{\rm BBH,max} - q_{\rm BBH,min}}.
    \label{eq: fmrr}
\end{equation}
This is shown in Figure~\ref{fig:q strong model} as a histogram in the top-right panel. 
The $f_{\rm MRR}$ posterior is bimodal, with a sharp spike near $f_{\rm MRR} \sim 0$ and a broader mode around $f_{\rm MRR} \sim 0.3$, indicating that even if MRR systems are present in the $10\,\Msun$ peak, they are likely a minority of the subpopulation.

The dominant non-MRR branch of the posterior accounts for $\macros[no_mrr_posterior_percent]\%$ of the total hyper-posterior. 
Hence, overall we find little support for MRR systems in the $10\,\Msun$ peak. 
This is consistent with the prediction of \citet{GINA:inprep} that MRR systems should preferentially populate higher masses. 
The constraint $\zeta_{\rm eff} < \zetaMRR$ also implicitly excludes $q_{\rm obs} = 1$. 
Other analyses \citep[e.g.,][]{LIGO:GWTC4Pop, Godfrey2023,Banagiri2025} have found that the mass ratio distribution of the 10 \Msun peak may prefer slightly unequal mass systems, which could be consistent with this finding. 

Lastly, the bottom-right panel of Figure~\ref{fig:q strong model} shows samples from the joint hyper-posterior on $\zeta_{\rm eff}$ and $\beta_{\rm acc}$. The two parameters appear mildly correlated, with smaller $\zeta_{\rm eff}$ favouring smaller $\beta_{\rm acc}$.
The marginal posteriors yield $\zeta_{\rm eff} = \CIPlusMinus{\macros[zeta_eff]}$ and $\beta_{\rm acc} = \CIPlusMinus{\macros[beta_acc]}$, both within astrophysically plausible ranges and corresponding to low-to-moderate mass transfer efficiency and fairly stable mass transfer. We caution that our SMT mass ratio model is a simplified description of the channel, and the inferred $\zeta_{\rm eff}$ and $\beta_{\rm acc}$ should be read as illustrative rather than as precise measurements. 
The contour lines in the bottom-right panel of Figure~\ref{fig:q strong model} show the joint $\zeta_{\rm eff}$-$\beta_{\rm acc}$ prior, divided into three of the four parameter regimes: no MRR systems (pink), a mix dominated by non-MRR (purple), and a mix dominated by MRR (gray). For the fixed $f_{\rm SN,a}$, $f_{\rm SN,b}$, $f_{\rm core}$ values used in our analysis, there is no prior support for a completely MRR population. We find no posterior support for reversed systems dominating over non-reversed systems.
%

%% file: conclusion.tex
\section{Discussion and Conclusion}\label{sec:conclusion}
In this work we have shown how `mass-ratio reversal' during binary evolution (i.e., cases where the initially less massive binary component creates the more massive compact remnant) can produce distinct mass ratio distributions, even within a single formation channel.
The shape of the mass ratio distributions is set by the fraction of the population that has undergone reversal, which in turn can be linked back to physical parameters of the binary mass transfer, such as the mass transfer efficiency $\betaacc$ and mass transfer stability parameters like the effective response to mass loss $\zetaeff$ (Figure \ref{fig: analytical q}).
Specifically, we show how the SMT mass-ratio model naturally splits into a non-reversed and a reversed subpopulation whose relative shapes depend on the SMT channel parameters. 
Across a substantial portion of physically plausible parameter space (including the regime we find to be most relevant to the $\sim 10\,\Msun$ peak), the non-reversed subpopulation is broad and can extend to low $q$, while the reversed subpopulation is narrower and concentrated near $q = 1$.

As a proof of concept, we develop a strongly parametrized model for the stable mass transfer channel (Section~\ref{ss:smt q model}) that maps the observed mass ratio distribution of merging black holes onto these binary evolution parameters. 
We applied this model to the low-mass peak in the BBH primary mass distribution, isolating systems in this peak following the approach of \citet{Godfrey2023}: using a Gaussian peak component for the primary mass with a separate mass ratio distribution, and a B-spline background capturing the rest of the population (Figure \ref{fig:primary masses}). These results highlight how data-driven models can act as "garbage collectors" in targeted studies of single formation channels, where the channel of interest may not account for the entire BBH population.

We find that the systems in the low-mass peak are most consistent with a population that is \textit{not} mass-ratio reversed, with a moderate-to-low mass transfer efficiency $\betaacc \sim 0.3$ and an effective response to mass loss $\zetaeff \approx 6.5$, corresponding to fairly stable mass transfer (Figure \ref{fig:q strong model}). 
Using population synthesis simulations of stable mass transfer, \citet{Broekgaarden:2022nst} find similar trends in the fraction of MRR systems $f_{\rm MRR}$ for different accretion efficiencies $\beta_{\rm acc}$; when $\beta_{\rm acc} = 0.25$, they find the fraction of MRR systems is near zero, while larger $\beta_{\rm acc}$ values lead to $f_{\rm MRR} = 0.2-0.6$. Though we don't consider systems that form a common envelope, i.e. where the second MT phase is unstable, \citet{Zevin:2022wrw} find the fraction of MRR systems for all MT scenarios in their rapid population synthesis models to increase as the accretion efficiency $\beta_{\rm acc}$ increases. \citet{Mould:2022xeu} and \citet{Adamcewicz:2023szp} use spin measurements to estimate the fraction of MRR systems in GWTC-3.0, as tidal spin-up of the initially less-massive star paired with MRR will lead to a spinning primary BH, rather than a spinning secondary BH. \citet{Mould:2022xeu} find the fraction of MRR systems to be $f_{\rm MRR} < 0.32$, while \citet{Adamcewicz:2023szp} finds a larger fraction $f_{\rm MRR} < 0.88$. Our analysis uses GWTC-4.0 and only considers the $10 \Msun$ peak as coming from isolated binary evolution, so the lack of evidence we find for these systems being mass-ratio reversed could indicate that MRR systems exist at larger primary masses in the catalog.
We stress that our strongly parametrized model is a gross simplification of the SMT channel: it treats $\betaacc$ and $\zetaeff$ as single population-wide values, when in reality these likely vary from system to system, and it does not capture all of the relevant physics, such as delayed dynamical instability.
The inferred values should therefore be read as illustrative rather than as precise measurements. Our goal is not to pin down the binary physics from current data, but to demonstrate that the mass ratio distribution \textit{can} be used to do so, and that mass ratio reversal in particular is a promising target to look for.

This work represents a first step/proof of concept and we envision several natural extensions and followup studies.
First, the inference could be extended to allow event-specific values of $\betaacc$ and $\zetaeff$, rather than treating them as single population-wide parameters.
Second, the SMT channel predicts both a non-reversed and a mass-ratio-reversed sub-population, which may occupy different mass ranges and exhibit different mass ratio distributions \citep{GINA:inprep}. 
A more flexible model would allow these two components to be coupled to separate primary mass B-Spline distributions, making it possible to test whether the data support a mass-ratio-reversed population at higher masses. This analysis is currently complicated by the order-of-magnitude lower merger rate in this regime, which makes it difficult to cleanly separate a peak from the background.
Third, the analytical model could be replaced by a surrogate trained on detailed binary population synthesis simulations, for example via normalizing flows \citep{Wong:2020ise, Colloms:2025hib, Plunkett:2025mjr}, so that the channel prediction can be fit alongside the flexible B-spline components without the sharp edges that limit our current implementation.
With GWTC-5.0 and future catalogs, the hierarchical constraints on individual mass ratios will tighten, and we expect the mass ratio distribution to become an increasingly powerful tool for identifying sub-populations and constraining the physics of the formation channels that shape the observed BBH population.